\documentclass[a4paper,twocolumn]{article}
\usepackage[english]{babel}
\usepackage[utf8]{inputenc}
\usepackage{amsmath}
\usepackage{mathrsfs}
\usepackage{graphicx}
\usepackage{mathtools}
\usepackage{rotating}
\usepackage{booktabs}
\usepackage{siunitx}
\usepackage{multirow}
\usepackage{subcaption}
\usepackage{hyperref}
\usepackage{xcolor}
\usepackage{lineno}
\usepackage{float}
\usepackage{placeins}
\usepackage{authblk}
\usepackage{abstract}
\hypersetup{colorlinks=true,linkcolor=black,citecolor=black}

\title{Yellow diode-pumped lasing of femtosecond-laser-written Dy,Tb:LiLuF\textsubscript{4} waveguide} 
\author[1]{Davide Baiocco}
\author[2,3]{Ignacio Lopez-Quintas} 
\author[2,3]{Javier R. Vázquez de Aldana} 
\author[4]{Alessandro di Maggio}
\author[4]{Fabio Pozzi}
\author[1]{Mauro Tonelli} 
\author[1]{Alessandro Tredicucci}
\affil[1]{Dipartimento di Fisica, Università di Pisa, Largo Bruno Pontecorvo 3, 56127 Pisa Italy}
\affil[2]{Grupo de Investigación en Aplicaciones del Láser y Fotónica, Universidad de Salamanca, Pl. La Merced SN.,Salamanca, 37008, Spain}
\affil[3]{Unidad de Excelencia en Luz y Materia Estructurada (LUMES), Universidad de Salamanca, Spain}
\affil[4]{Lumibird Photonics Italia Srl,Via G. Schiaparelli, 12, Torino, 10148, Italy}

\date{December 10, 2024}
\begin{document}
\twocolumn[{%
  \begin{@twocolumnfalse}
    \maketitle
    \begin{abstract}

In this article we report the fabrication of a diode-pumped Dy,Tb:LiLuF\textsubscript{4} waveguide laser operating in the yellow region of the visible spectrum.
The circular depressed-cladding waveguides have been fabricated by direct femtosecond laser writing, and showed propagation losses as low as \SI{0.07}{\deci\bel/\centi\meter}.
By employing these structures, we obtain a maximum output power of \SI{86}{\milli\watt} at \SI{574}{\nano\meter} from a \SI{60}{\micro\meter} diameter waveguide, and a highest slope efficiency of 19\% from a \SI{80}{\micro\meter} diameter depressed cladding waveguide.
 In addition, we demonstrate lasing at \SI{574}{\nano\meter} from a half-ring surface waveguide, with a maximum output power of \SI{12}{\milli\watt}.
 Moreover, we also obtained dual wavelength operation at 568-574\,\si{\nano\meter}, with a maximum output power of \SI{15}{\milli\watt}, and stable lasing at \SI{578}{\nano\meter}, with an output power of \SI{100}{\milli\watt}. The latter wavelength corresponds to the main transition of the atomic clock based on the neutral ytterbium atom.
 To the best of the authors' knowledge, this is the first demonstration of a yellow waveguide laser based on Dy-doped materials.
\end{abstract}
  \end{@twocolumnfalse}
}]

\section{Introduction}
The lack of semiconductor lasers emitting in the yellow band led to research for alternative compact sources. 
In fact, coherent sources emitting in the yellow band of the spectrum are employed for medical applications~\cite{YELLOWMEDICAL,YELLOWMEDICAL2,YELLOWMEDICAL3}, as well as in the scientific and industrial field~\cite{GIALLOREVIEW}.
Moreover, the {$^1\mathrm{S}_{0}\rightarrow^3\!\!\mathrm{P}_{0}$} atomic clock transition of neutral Yb is located at \SI{578}{\nano\meter}~\cite{YBCLOCK}.
The sources employed for these clocks are typically based on sum frequency generation~\cite{YBCLOCK} or second harmonic generation~\cite{YBSHG}. Nevertheless, optical setups involving nonlinear optics are more complex, while a source directly emitting in the yellow will be easier to stabilize and will result more compact and lightweight, both crucial aspects for devices designed for the application in the aero-spatial field.

Among rare earth-based lasers, dysprosium demonstrated the possibility of laser operation in the 560-580\,\si{\nano\meter} band. In addition, it shows an absorption transition around \SI{450}{\nano\meter}, within the emission band of InGaN-based laser diodes, today commercially available with watt-level output power.
The main limitation of  Dy\textsuperscript{3+} is the presence of a spin flip, both in the absorption and in the laser transition, leading to absorption and emission cross sections two orders of magnitude lower when compared to neodymium or praseodymium~\cite{BLUE}. Moreover, the doping is limited to a few percent from detrimental upconversion and cross-relaxation processes, which reduce the mean lifetime of the upper laser levels.
 This constrains the active medium to be in the the 2-3\,\si{\centi\meter} length range~\cite{GIALLOBOLOGNESI,DYHUBER,DYZHANGSCALING,DYWANG,DYZNWO}, reflecting in a poor overlap between the laser mode inside the cavity and the pump beam, with consequent low efficiencies and high laser thresholds. 
An ideal matrix to host dysprosium is given by fluoride crystals, due to their low crystal field that prevents the dipole-allowed excited state absorption to the 4f\textsuperscript{8}5d levels.
Moreover, fluorides are not hygroscopic and are chemically stable, allowing device fabrication and operation in severe conditions.

Waveguide lasers are a possible solution to the above problem, due to the possibility of confining both the pump and the laser mode for all the length of the crystal. A technique that has been extensively employed for the fabrication of waveguide lasers is direct femtosecond writing~\cite{DAVISFEMTO}. By irradiating transparent dielectrics with laser pulses with a temporal duration of few tens of femtoseconds, it is possible to permanently change the refractive index of the irradiated area, allowing the fabrication of cladding waveguides in glass and crystals~\cite{BOOKFEMTOSECOND}.
Infrared femtosecond-laser-written waveguide lasers have been demonstrated, exploiting for example Nd\textsuperscript{3+}~\cite{NDYAG} or Yb\textsuperscript{3+}~\cite{CALMANOYB}, while visible Pr-based lasers have been fabricated only in the last years~\cite{CALMANOMGSRALO,OLBAIOCCO,PHOTONICSBAIOCCO,BAIOCCOOE,BAIOCCOPRIR}.
On the contrary, Dy-based waveguide laser have never been reported, due to the necessity of high quality materials and low loss waveguides in the  visible range.

In this work we report the first, to the best of the authors' knowledge, fabrication of a Dy,Tb:LiLuF\textsubscript{4} waveguide laser working in the yellow region of the visible spectrum. We demonstrated laser operation at \SI{568}{\nano\meter}, \SI{574}{\nano\meter}, and \SI{578}{\nano\meter}, showing values of output power of the order of \SI{100}{\milli\watt} and maximum slope efficiencies of 19\%. 
\section{Crystal growth and spectroscopic analysis}
We selected to work with LiLuF\textsubscript{4} (LLF) codoped with Dy\textsuperscript{3+} and Tb\textsuperscript{3+} to optimize the performance of the laser.
The choice of such crystal matrix is due to its better thermomechanical properties with respect to LiYF\textsubscript{4}~\cite{PROPLLF}. These properties were tested with Pr-doped LLF~\cite{BAIOCCOTHERMAL}.
The chosen doping was 4\% Dy\textsuperscript{3+} and 1\% Tb\textsuperscript{3+}. The value of Dy\textsuperscript{3+} doping was chosen to maximize the absorption, avoiding the reduction of the mean lifetime of the $^4\mathrm{F}_{9/2}$ upper laser manifold due to nonradiative decay channels. The 1\% codoping with Tb is added to reduce the mean lifetime of the $^6\mathrm{H}_{13/2}$ lower laser level, due to the energy transfer channel to the $^7\mathrm{F}_{4}$ manifold of Tb\textsuperscript{3+}. This leads to better performance of yellow lasers~\cite{GIALLOBOLOGNESI}. With this codoping, the mean lifetime of $^6\mathrm{H}_{13/2}$  was found to be \SI{58}{\micro\second } instead of about \SI{300}{\micro\second} measured on 4\% doped Dy:LLF. On the contrary, the $^4\mathrm{F}_{9/2}$ manifold is almost unaffected by the codoping, demonstrating a reduction of the mean lifetime of about \SI{20}{\micro\second} and a corresponding mean lifetime of about \SI{1.33}{\milli\second}~\cite{GIALLOBOLOGNESI}.

\begin{figure}[!htb]
\centering
\includegraphics[width=\columnwidth]{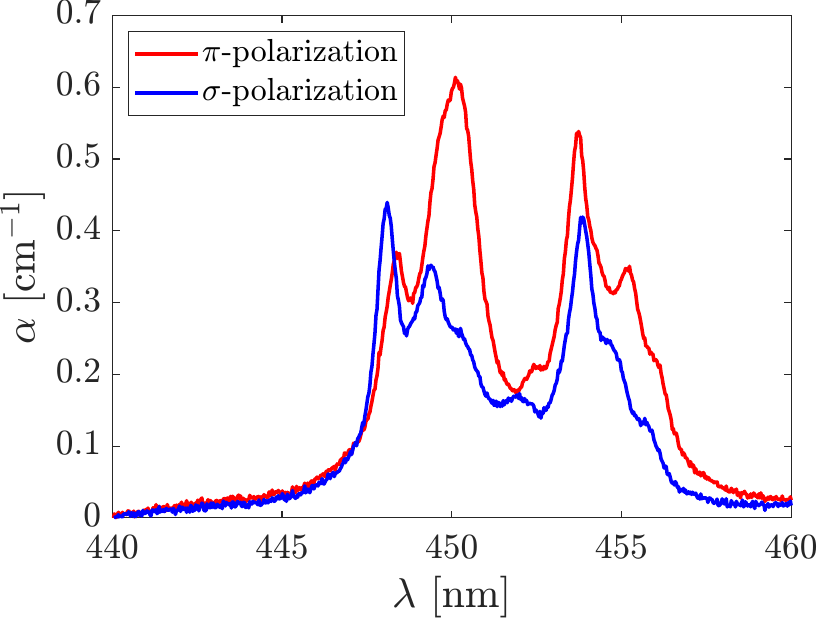}
\caption{Polarized absorption spectra of Dy,Tb:LLF. Resolution of \SI{0.09}{\nano\meter}.}
\label{ABS}
\end{figure}

Dy,Tb:LLF was grown by Czochralski technique from high purity powders of the constituents, LiF, LuF\textsubscript{3}, DyF\textsubscript{3}, and TbF\textsubscript{3}. After the growth, the boule was oriented by x-ray backscattering technique to cut oriented samples.
We performed absorption spectroscopy with a CARY 5000 spectrophotometer, reported in figure~\ref{ABS}.
The main absorption peak related to the $^6\mathrm{H}_{15/2}\rightarrow^4\!\!\mathrm{F}_{9/2}$  is located at \SI{450}{\nano\meter} for $\pi$-polarization, with an absorption coefficient of about \SI{0.6}{\centi\meter^{-1}}.
Due to the measured value of the absorption coefficient, we decided to cut a  ${\SI{4}{\milli\meter} \mathrm{(a)}\times\SI{4}{\milli\meter}\mathrm{(c)}\times\SI{19}{\milli\meter} \mathrm{(a)} }$ sample, in order to write \SI{19}{\milli\meter}-long waveguides. 
We chose a crystal length of about \SI{20}{\milli\meter}, with a final length of \SI{19}{\milli\meter} after the polishing,  to ensure a minimum single-pass absorption of the pump beam of about 40\%.

For the waveguides inscription, a Ti:sapphire femtosecond laser (Spitfire, Spectra Physics) was used. The output beam has pulse duration of \SI{60}{\femto\second}, a central wavelength of \SI{800}{\nano\meter} and operates at a repetition rate of \SI{5}{\kilo\hertz}. The writing procedure was based on a sample scanning approach, where the sample is place in a 3-axes motorized state, and the beam is focused by a 40X microscope objective lens. The fabricated waveguides consisted of depressed-index cladding structures~\cite{NDYAG} with optimized circular geometry and different core diameters, along the \SI{19}{\milli\meter} (a) axis. The laser polarization was kept perpendicular to the waveguide direction, along the (a) axis. A pulse energy of \SI{92}{\nano\joule} and a scanning velocity of \SI{1.2}{\milli\meter/\second} was found to be the optimum fabrication condition for the buried waveguides, and \SI{76}{\nano\joule} with a scanning velocity of \SI{0.6}{\milli\meter/\second} for the surface (half-ring) structures.

\begin{figure}[!htb]
    \centering
\begin{subfigure}[t]{0.49\columnwidth}
        \includegraphics[width=\textwidth]{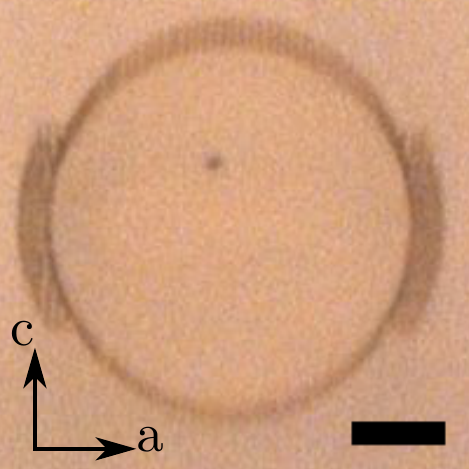}
    \end{subfigure}
    \hfill
  \begin{subfigure}[t]{0.49\columnwidth}
    \includegraphics[width=\textwidth]{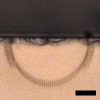}
    \end{subfigure}
\caption{Microscope image of WG2 (left, \SI{80}{\micro\meter} diameter) and WG3 (right, \SI{80}{\micro\meter} diameter, half waveguide) waveguide cross sections. The crystallographic axes are common to both pictures. Scale bars correspond to \SI{20}{\micro\meter}.}
\label{GUIDE}
\end{figure}

The best results here reported concern three waveguides. Two of them are ear-like waveguides (cladding optimization for a better light confinement) with diameters of \SI{60}{\micro\meter} (WG1) and \SI{80}{\micro\meter} (WG2), while the third is a half ear like waveguide written in direct contact with the crystal surface in order to use the air-crystal interface as part of the waveguide itself. This third waveguide has a diameter of \SI{80}{\micro\meter} (WG3). Pictures showing the two different geometries are reported in figure \ref{GUIDE}.

To perform all the experiments, the sample containing the waveguides was inserted in a custom-made copper holder. Between the holder and the crystal, an indium foil was placed to improve heat dissipation from the crystal. The holder was thermally connected with a thermoelectric cooler driven by a custom made temperature stabilizer. 
Waveguides written in LLF with similar geometries demonstrated propagation losses between \SI{0.1}{\deci\bel/\centi\meter} and \SI{0.2}{\deci\bel/\centi\meter}~\cite{OLBAIOCCO,PHOTONICSBAIOCCO,BAIOCCOOE}. The value of the current propagation losses in this case will be estimated from the laser experiments.
As pump source we chose an InGaN-based laser diode, manufactured and internally collimated by Lumibird Photonics Italia. 
We tuned this source to emit at \SI{447}{\nano\meter}. This value was a compromise between the maximum output power available from the diode and the crystal absorption efficiency. The pump beam was controlled with a half-wavelength plate and a polarizing beam splitter to vary the beam power, while a second half-wavelength plate was used to adjust the beam polarization. The maximum pump power available in front of the laser cavity is about \SI{2.7}{\watt}.
For $\pi$-polarized light, the single pass absorption efficiency is 40\%. 
Before executing the laser experiments, we measured the coupling efficiency of the waveguides. To do that, we performed coupling and transmission measurements testing various lenses as coupling lens and using a \SI{20}{\milli\meter} aspherical lens to collect the beam exiting from the waveguide. Both lenses were mounted on two independent THORLABS MBT616D/M precision positioning stages, to optimize the coupling and the collection of the beams in the waveguide.

The best results have been observed using an aspherical lens with \SI{30}{\milli\meter} focal length as the coupling lens. WG1 demonstrated a coupling efficiency of about 75\%, WG2 of 77\% and WG3 of 58\%.
To estimate this value, we neglected the contribution of the propagation losses with respect to the absorption of Dy\textsuperscript{3+} ions.
This means assuming the following relation between the transmission efficiency ($\eta_T$), the coupling efficiency ($\eta_C$) and the material absorption:
\begin{equation}
    \eta_T=\frac{P_{o}}{P_{inc}}=F_{IN}F_{OUT}\eta_Ce^{-\alpha L}
    \label{EQPERDITE}
\end{equation}
where $\eta_T$ is the ratio between power exiting from the waveguide ($P_{O}$) and the incident power ($P_{INC}$), $F_{IN}$ and $F_{OUT} $ are the Fresnel transmission efficiencies at the two air-crystal interfaces, $\eta_C$ is the coupling efficiency, L is the crystal length, and $\alpha\approx\SI{0.3}{\centi\meter^{-1}}$ is the absorption of the crystal matrix at the pump wavelength.

No saturation of pump absorption was observed in the waveguides studied. We also observed the intensity profile of the waveguides. To do that, we placed a CCD camera in the conjugate plane of the crystal facet, to observe the intensity profile of the confined mode. The waveguides resulted multimodal at the pump wavelength, as observed in previous experiments with ear like waveguides in LLF~\cite{BAIOCCOOE}.
Since it is possible to estimate the coupling efficiency, slope efficiencies and laser threshold will be given as a function of the pump power coupled in the waveguide.
\section{Laser experiments}
To build the laser cavity, two plane mirrors were approached at a short distance ($<\SI{0.1}{\milli\meter}$) from the crystal facets. The mirrors were mounted in two custom-made mirror holders composed by a three-axes translation stage and a two axes tilting stage, in order to independently optimize the position and tilt of each mirror.
A total of three mirrors were employed for all the laser experiments, one with the function of input coupler (IC) and two different output couplers with different values of extraction in the yellow band. The transmission spectra of each mirror was measured with the CARY5000 spectrophotometer and are reported in table~\ref{SPECCHI}.

\begin{figure}[!htb]
\centering
\includegraphics[width=\columnwidth]{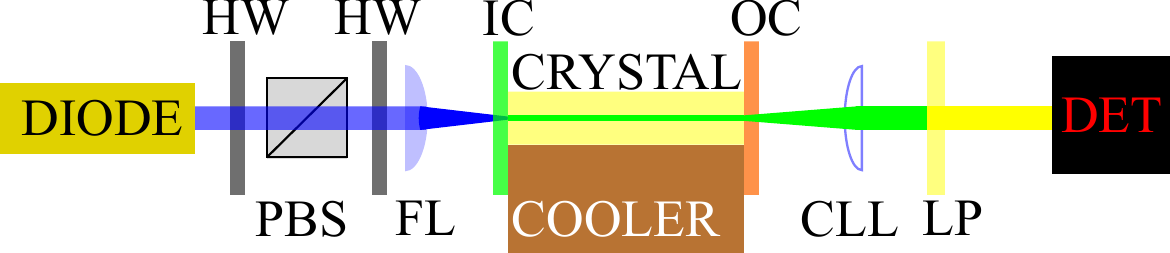}
\caption{Schematics of the whole setup. DIODE is the collimated pump diode, HW are the two half-wavelength plates, PBS is the polarizing beam splitter, FL is the focusing lens, IC is the input coupler, CRYSTAL is the Dy,Tb :LLF crystal, OC is the output coupler, CLL is the collection lens, LP is the long pass dielectric filter, and DET represents the various instruments employed to analyze the laser beam.}
\label{SCHEMA}
\end{figure}

\begin{table}[!htb]
    \centering
    \begin{tabular}{cccccc}\hline
        \multirow{2}{*}{Mirror}& \multirow{2}{*}{Use}& \multicolumn{4}{c}{Transmittance [\%] at $\lambda$ [\si{\nano\meter}]}\\
        &&447&568&574&578\\\hline
        1&IC&91&0.006&0.006&0.006\\\hline
        2&OC&91&0.6&0.6&0.7\\\hline
        3&OC&6.0&4.5&5.0&6.0\\\hline
        \end{tabular}
    \caption{Summary of the characteristics of the employed mirrors. IC means input coupler while OC stands for output coupler.}
    \label{SPECCHI}
\end{table}
The beam exiting from the waveguide was collected with the \SI{20}{\milli\meter} collection lens and the residual pump power was removed with a \SI{510}{\nano\meter} long-pass dielectric filter.
A schematic of the whole system is reported in figure~\ref{SCHEMA}.

\begin{figure}[!htb]
\centering
\includegraphics[width=\columnwidth]{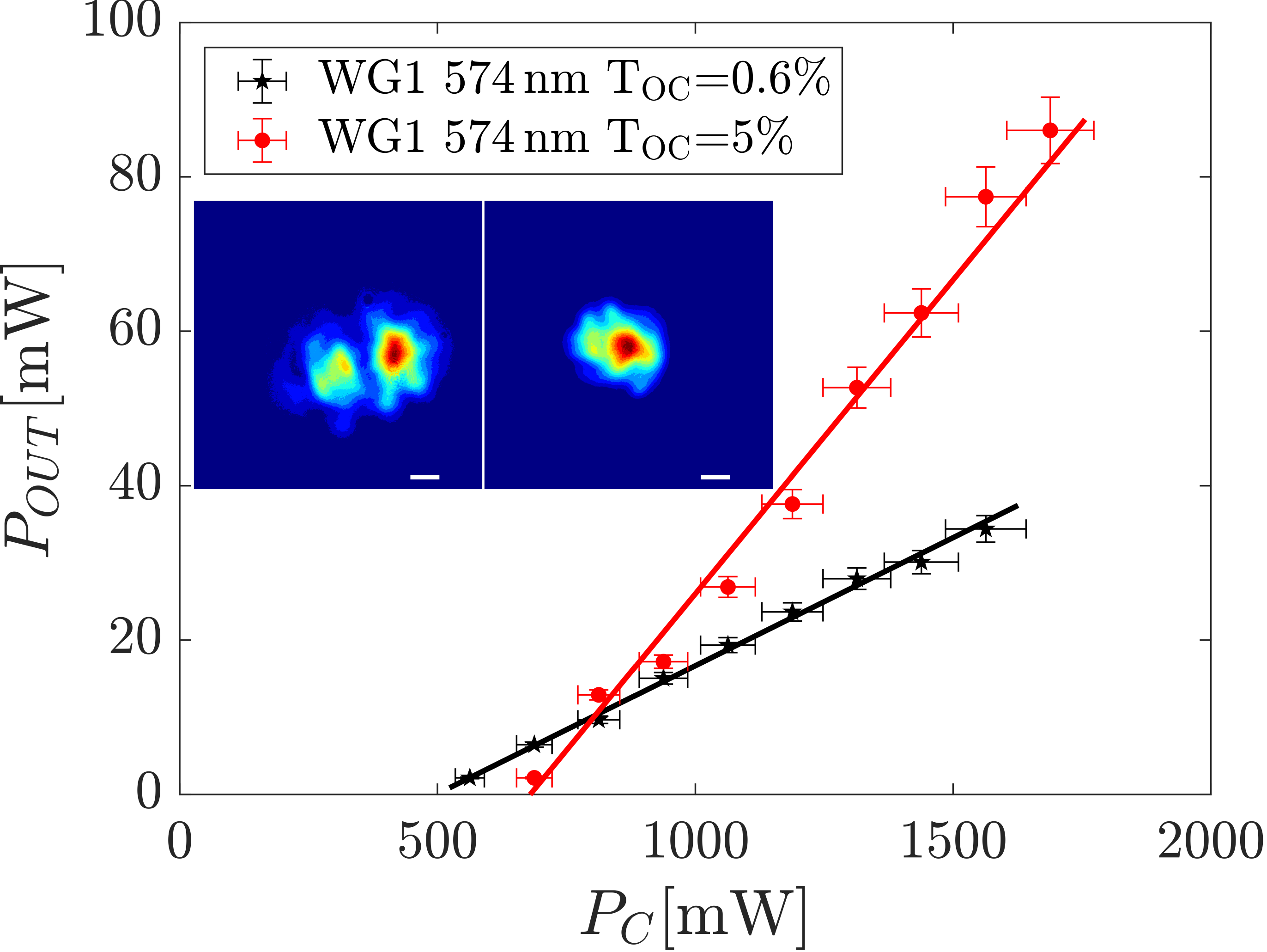}
\caption{Data, best fit, and intensity profiles for the laser operation of WG1 at \SI{574}{\nano\meter}. The abscissa reports the pump power coupled in the waveguide.} The left inset corresponds to the mode achieved with mirror 2 as OC, while the right one was obtained with mirror 3. Both profiles were acquired at maximum pump power. Scale bars correspond to \SI{10}{\micro\meter}.
\label{SLOPEWG1C}
\end{figure}

The best laser performances were achieved with WG1. By employing mirror 1 as IC and mirror 2 as OC, the laser demonstrated operation at \SI{574}{\nano\meter} with a maximum output power of \SI{35}{\milli\watt}, a slope efficiency of 3\% and a threshold power of \SI{550}{\milli\watt}. Substituting mirror 2 with mirror 3, the laser demonstrated a maximum output power of \SI{86}{\milli\watt}, a slope efficiency of 7\%, and a threshold power of \SI{680}{\milli\watt}. Data, best fit, and intensity profiles for the two extractions are reported in figure~\ref{SLOPEWG1C}.

\begin{figure}[!htb]
\centering
\includegraphics[width=\columnwidth]{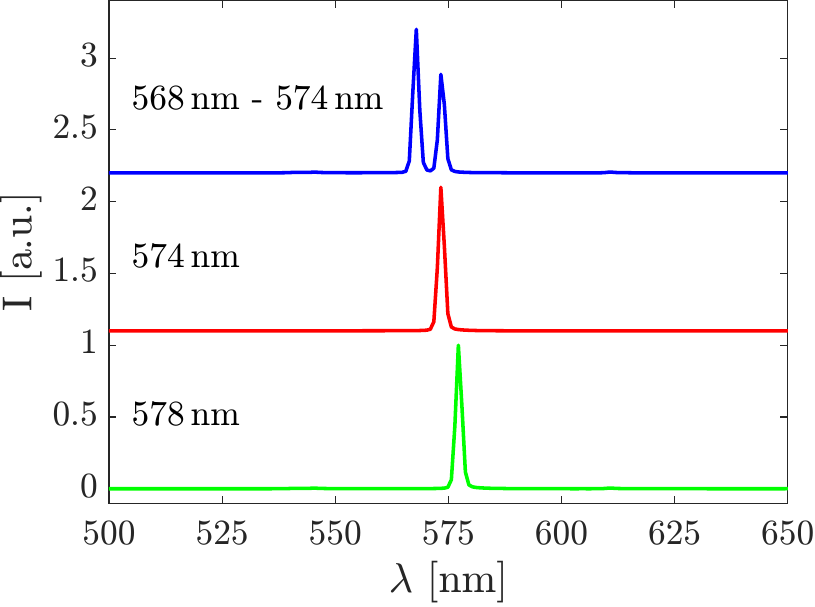}
\caption{Emission spectra of the single and dual-wavelength operation of the yellow DyTb-waveguide laser. Resolution of \SI{1}{\nano\meter}.}
\label{SPETTRI}
\end{figure}

The M\textsuperscript{2} of the laser emissions have been measured with a Coherent ModeMaster MM-2S, resulting in values between 4 and 5 for the two principal axes of the beam.
Moreover, by slightly tilting mirror 3, tuning of the laser was observed. In particular, a maximum output power of \SI{100}{\milli\watt} was observed at \SI{578}{\nano\meter} as well as dual wavelength operation at \SI{568}{\nano\meter}-\SI{574}{\nano\meter}, with a total output power of \SI{15}{\milli\watt}. 
The higher output power observed on the weaker emission line (\SI{578}{\nano\meter}) is due to the higher extraction of the OC at this wavelength (6\%) with respect to the one at the main line (5\%) and to the different overlap efficiency between the pump beam and the laser mode at different wavelengths inside the multimodal waveguide.
By lowering the pump power, only dual wavelength \SI{574}{\nano\meter}-\SI{578}{\nano\meter} operation was achieved, impeding the measurement of the slope efficiency at \SI{578}{\nano\meter}. Laser emission spectra, acquired with a QEpro spectrophotometer, are reported in figure~\ref{SPETTRI}.

\begin{figure}[!htb]
\centering
\includegraphics[width=\columnwidth]{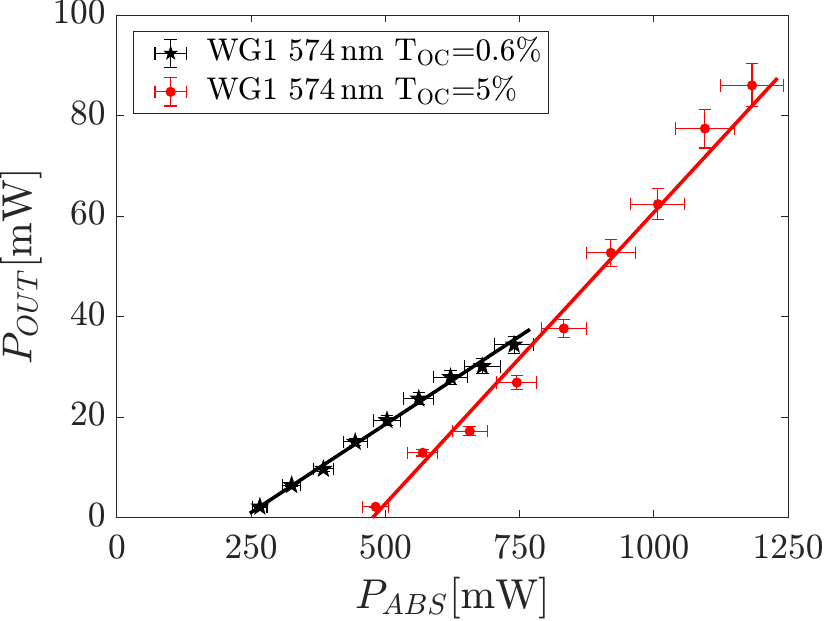}
\caption{Data and best fit for the laser operation of WG1 at \SI{574}{\nano\meter}, taking into account the multi-pass absorption efficiency. The abscissa reports the pump power absorbed by the active medium.} 
\label{SLOPEWG1ABS}
\end{figure}

Mirror 2 and 3 demonstrate a completely different behavior at the pump wavelength. In fact, mirror 2 reflects 9\% of the residual pump power, while mirror 3 reflects 94\%. Due to the geometry of the laser cavity, the residual pump power is reflected back in the waveguide, allowing a multi-pass pump scheme.
Owing to this effect, to better compare the results achieved with the two different extractions it is necessary to evaluate the pump power effectively absorbed by the waveguide core.
It is possible to perform this calculation since the active medium did not show saturation effects.
To do that, we assumed a complete reinsertion of the pump exiting from the waveguide at each reflection at the mirrors. This calculation leads to better estimates of the laser performance.
By evaluating slope efficiency and threshold power as a function of the absorbed power, the laser employing mirror 2 demonstrated a slope efficiency of 7\% and a threshold power of \SI{230}{\milli\watt}, while a slope efficiency of 12\% and a threshold power of \SI{480}{\milli\watt} were observed employing mirror 3. The slope efficiencies demonstrated are comparable to those achieved in bulk lasers, while demonstrating lower thresholds, highlighting the better overlap between the pump and the laser mode in the waveguide core with respect to bulk crystals~\cite{GIALLOBOLOGNESI,DYWANG,DYZHANGSCALING,DYHUBER,DYZNWO}. 
Data and best fit are reported in figure~\ref{SLOPEWG1ABS}.

\begin{figure}[!htb]
\centering
\includegraphics[width=\columnwidth]{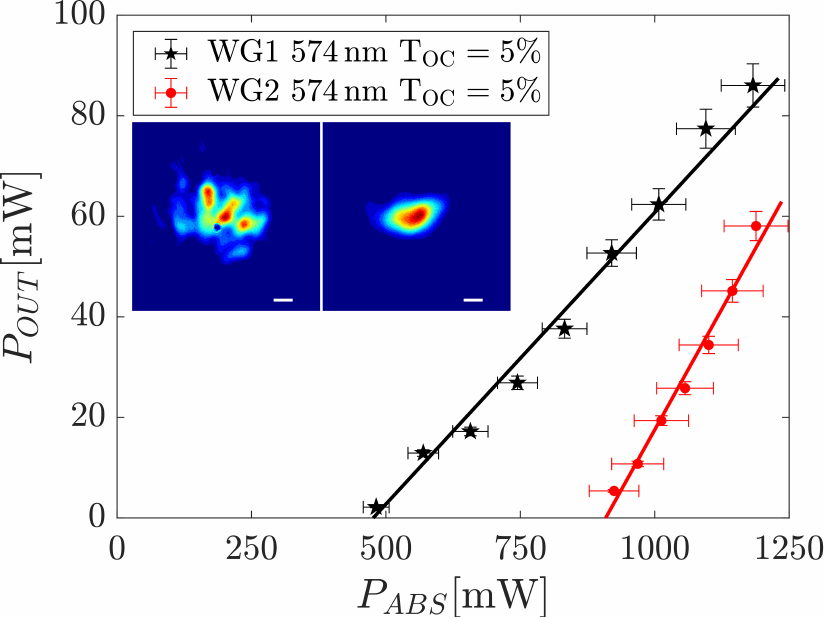}
\caption{Data and best fit for the laser operation of WG2 at \SI{574}{\nano\meter}. The data from WG1 are also reported for comparison. The left inset shows the emission profile of WG2, while the right inset reports the WG3 one. Both were acquired at maximum pump power. The scale bars correspond to \SI{10}{\micro\meter}.}
\label{WG2}
\end{figure}

After that, we tested WG2 employing mirrors 1 and 3 to build the laser cavity.
With this configuration, we observed laser operation at \SI{574}{\nano\meter}, with a maximum output power of \SI{58}{\milli\watt}, a slope efficiency with respect to the absorbed power of 19\%, and a threshold power of \SI{900}{\milli\watt} absorbed by the active medium.
From the values of the threshold power, we performed a Findlay-Clay analysis to estimate the propagation losses. The obtained value is \SI{0.07}{\deci\bel/\centi\meter}. Such value is comparable with those measured in femtosecond-laser-written waveguides fabricated in fluorides~\cite{OLBAIOCCO,BAIOCCOOE}, and justifies the approximations made to estimate the coupling losses (see Eq.~\ref{EQPERDITE}).
Data, best fit and intensity profile are shown in figure~\ref{WG2}.
The ratio between the threshold power observed with WG2 and WG1, employing the same cavity configuration, is about 1.9 and it can be attributed to the different values of the core transverse area, whose ratio is equal to 1.8. The slope efficiency is higher than WG1 and is the highest among all diode-pumped lasers based on Dy-doped crystals~\cite{GIALLOBOLOGNESI,DYWANG,DYZHANGSCALING}, result achieved exploiting the better overlap between pump and laser mode within the waveguide.
In the end, we achieved stable lasing also from WG3, that demonstrates \SI{12}{\milli\watt} output power at maximum pump power ({$\mathrm{P_{ABS}}=\SI{900}{\milli\watt}$}), operating at \SI{574}{\nano\meter} with mirror 3 as OC. The achieved output power was too low to allow the determination of the slope efficiency. Nevertheless, since the crystal facet is part of the core-cladding interface for all the waveguide length, this waveguide geometry can be a promising platform for the interaction of the laser mode with saturable absorbers or dielectric structures to realize a monolithic laser cavity.
The intensity profile is reported in the inset of figure~\ref{WG2}.
All laser emissions are $\pi$-polarized. The laser results are summarized in table~\ref{LASER}.

\begin{table*}[!htb]
    \centering
    \begin{tabular}{cccccccc}\hline
         WG&$\lambda$[nm]&$\mathrm{P_{OUT}}$  &$\mathrm{P_{THR}} $&$\eta_{ABS}$&$T_{OC}$ &$\mathrm{M^2_x}\times\mathrm{M^2_y}$\\\hline\hline
        \multirow{4}{*}{WG1}& {574}&86&480&12&5&$5\times4$    \\ 
         &  {574}&35&230&7&0.6&$5\times5$\\ 
    &   {578}&100&-&-&6&$4\times4$   \\   
         &  {568-574}&15&-&-&4.5-5&-\\\hline
     {WG2}&574&58&900&19&5&$4\times4$\\\hline 
            WG3 &  574&12&-&-&5&$4\times3$\\\hline

   \end{tabular}
    \caption{Summary of the laser results with the yellow Dy,Tb:LLF waveguide lasers. P\textsubscript{THR} is the absorbed threshold power. P\textsubscript{THR} and P\textsubscript{OUT} are measured in milliwatt, while $\eta_{ABS}$ and T\textsubscript{OC} are expressed as a percentage.}
    \label{LASER}
\end{table*}
\section{Conclusion}
In conclusion, we demonstrated the first laser operation of a Dy-based waveguide laser working in the yellow band of the visible spectrum. We achieved a maximum output power of \SI{86}{\milli\watt} at \SI{574}{\nano\meter} and \SI{100}{\milli\watt} at \SI{578}{\nano\meter}. Moreover, we demonstrated the possibility of laser operation at \SI{568}{\nano\meter}, with a total output power in dual wavelength regime of \SI{15}{\milli\watt}.
The maximum slope efficiency observed is 19\%, which is the highest value for crystal based yellow sources.
This geometry can lead to the fabrication of compact monolithic sources for metrological aerospace-oriented devices.
In the end, we also proved the possibility of laser operation of a half-waveguide written at the crystal surface, delivering a maximum output power of \SI{12}{\milli\watt} at \SI{574}{\nano\meter}. This geometry allows the direct interaction between the laser mode and dielectric structures fabricated on the crystal surface, facilitating the fabrication of sources based on microresonators. The output power is only limited by the available pump power and absorption efficiency at the pump wavelength chosen.
A power scaling of the device can be carried out by pumping at resonance, corresponding to a wavelength of \SI{450}{\nano\meter}, since the crystal matrix and the waveguide cladding have been shown to be able to withstand thermally severe conditions~\cite{BAIOCCOTHERMAL}.
\FloatBarrier
\section*{Declaration of competing interest}
The authors declare no conflict of interest.
\section*{Funding}
 Ministerio de Ciencia, Innovación y Universidades (PID2020-119818 and PID2023-149836NB);
 \section*{Data Availability Statement}
Data underlying the results presented in this paper are not publicly available at this time but may be obtained from the corresponding author upon reasonable request.
\section*{Author Contribution}
Davide Baiocco: Investigation, Writing – original draft, Writing – review \& editing.
Ignacio Lopez-Quintas: Investigation, Waveguide design and fabrication, Writing – original draft, Writing – review \& editing.
Javier R. Vázquez de Aldana: Investigation, Waveguide design and fabrication, Writing – original draft, Writing – review \& editing.
Alessandro di Maggio:Investigation, Writing – original draft, Writing – review \& editing.
Fabio Pozzi:Investigation, Writing – original draft, Writing – review \& editing.
Mauro Tonelli: Investigation, Writing – original draft, Writing – review \& editing.
Alessandro Tredicucci: Investigation, Writing – original draft, Writing – review \& editing.
\bibliographystyle{IEEEtran}

\bibliography{biblio.bib}
\end{document}